 \documentclass[preprint2]{aastex}
\usepackage{epsfig}

\slugcomment{ Submitted to Publications of the Astronomical Society of
the Pacific: 2001 Jan 10, Accepted 2001 Jan 19}

\shorttitle{Leuschner Infrared Camera}
\shortauthors{Graham \& Treffers}

\begin{document}


\title{An Infrared Camera for Leuschner Observatory \\
and the Berkeley Undergraduate Astronomy Lab}


\author{James R. Graham and Richard R. Treffers}
\affil{Astronomy Department, University of California, Berkeley, CA 94720-3411}

\email{jrg, rtreffers@astron.berkeley.edu}




\begin{abstract}

We describe the design, fabrication, and operation of an infrared
camera which is in use at the 30-inch telescope of the Leuschner
Observatory.  The camera is based on a Rockwell PICNIC 256$\times$256
pixel HgCdTe array, which is sensitive from 0.9-2.5 $\mu$m.  The
primary purpose of this telescope is for undergraduate instruction.
The cost of the camera has been minimized by using commercial parts
whereever practical. The camera optics are based on a modified Offner
relay which forms a cold pupil where stray thermal radiation from the
telescope is baffled. A cold, six-position filter wheel is driven by a
cryogenic stepper motor, thus avoiding any mechanical feed throughs.
The array control and readout electronics are based on standard PC
cards; the only custom component is a simple interface card which
buffers the clocks and amplifies the analog signals from the array.

\vskip 0.1 in
\noindent
{\it Submitted to Publications of the Astronomical Society of the
Pacific: 2001 Jan 10, Accepted 2001 Jan 19}

\end{abstract}




\section{Introduction}

Many undergraduates enter universities and colleges with the ambition
of pursing several science courses, or even a science major.  Yet
significant numbers take only the minimum requirements because science
is often perceived as laborious, demanding, and irrelevant.  The
result is graduates whose lives and careers as citizens are not
enriched by an understanding of science or technology.  Since the
Astronomy Department is frequently an undergraduate's only contact
with science we have a tremendous responsibility as a provider of
undergraduate science instruction.  Unfortunately, the students who
are captivated by introductory astronomy often conclude that the
astronomy major consists of narrow technical courses designed only for
those intent on pursuing an academic career. These classes 
often do not
stress problem solving, critical thinking nor emphasize the
connections between broad areas of scientific knowledge. The Berkeley
undergraduate astronomy labs offer a potent antidote to conventional
lectures, and stimulate students to pursue the astronomy major.  We
describe a state-of-the-art infrared camera which is used at Berkeley
to inspire and instruct.

Prior to the construction of this camera the undergraduate lab
syllabus consisted of an optical and a radio class.  The labs use
theodolites, a CCD-equipped robotic observatory (Treffers et
al. 1992), and radio telescopes to provide the astronomical context to
discuss the statistical description of experimental data, including
the notions of errors, error propagation, and signal-to-noise ratio.
These resources also provide students with a practical forum in which
to explore signal processing techniques including power spectrum
estimation and convolution. The optical labs are traditionally offered
in the Fall semester. The Spring climate in Northern California is
unsuitable for optical observations, so a radio-astronomy lab based on
a roof-top 21-cm horn antenna and a two-element 12 GHz satellite-dish
interferometer is offered (Parthasarathy et al. 1998).

These experiments have significant computational demands that involve
least square fitting, precession, coordinate conversion, Doppler
correction, Fourier transforms, and image processing.  Students use a
cluster of Sun Ultra workstations and the IDL programming environment
for data reduction and analysis.  These machines provide, for example,
the data storage and computational power required for four to five
groups of students working simultaneously to turn the visibility data
from the interferometer into radio maps.

\subsection{Development of an infrared undergraduate laboratory}

To maintain the intellectual vitality of the undergraduate lab,
broaden the range of observing experiences we offer, and satisfy the
demand for increased enrollment we have developed a new lab program
based on the theme of infrared astronomy.  Although radio receivers
are commonplace, the notion of using radio technology for
astronomy remains exotic. Radio astronomy continues to exemplify
astronomy of the ``invisible universe'', and detecting astronomical
radio signals continues to intrigue and captivate students.  Perhaps,
because our eyes give us an intuitive sense of what it means to
``see'' at visible wavelengths, and because CCDs have become
commonplace in products such as digital cameras, the high-tech aspect
of the optical lab has lost some of its gloss.  
Modern CCDs have excellent cosmetic quality, and students
report that calibration of CCD images appears to consist of annoying
minor corrections. Faint stars and galaxies are readily observable in
raw CCD images.  In contrast, at infrared wavelengths the brightness
of the sky (e.g., 13 mag. per square arcsec at $K$-band), compounded
by detector flaws and artifacts, means that it is impossible to detect
interesting sources without sky-subtraction and flat-fielding.
Digging the infrared signal of optically-invisible dust-enshrouded
young stars out of noise can be presented as a challenging and
worthwhile goal (cf. \S \ref{example-observation}).

Thus, infrared astronomy is a natural next step for an undergraduate
observatory that is equipped with optical and radio facilities.
Moreover, infrared observations are becoming familiar to students from
introductory undergraduate text books which rely on observations from
{\it IRAS}, {\it ISO}, and and the {\it NICMOS} camera on the {\it
Hubble Space Telescope} to describe astrophysical phenomena such as
star formation.  We can expect interest to soar as {\it SOFIA} and
{\it SIRTF} become operational in the next few years.  Infrared
observations provide a unique context to discuss a broad range of
physical phenomena and astrophysical processes: black-body radiation,
solid state physics and the operation of photon detectors, telescope
optics, and absorption and emission by dust and molecules.

Because of the complexities associated with cryogenic optics needed
for operation at wavelengths longer than about 1.6 $\mu$m, a prototype
camera with a fixed $H$-filter located immediately in front of the
focal plane array was fielded for instructional purposes in the Fall of
1999. This version of the camera provided the opportunity to verify
the operation of the array readout electronics. A Mark II camera
with cold re-imaging optics and a cryogenic filter wheel was constructed
during Summer 2000, and used on a class conducted in the Fall 2000
semester.

\section{Design of the Infrared Camera}

Given the restricted budgets available for undergraduate instruction
the single most important factor influencing the design of the camera
is inevitably cost.  Traditionally, infrared cameras require expensive
custom components---cryogenic vacuum vessels, cryogenic mechanisms,
optics, and array control and readout electronics.  We have controlled
costs by adopting a design which relies almost entirely on commercial
parts.

The heart of the camera is the infrared sensor.  Due to the generous
support of the National Science Foundation and provision of matching
funds by Rockwell International we were able to procure an engineering
grade $256 \times 256$ HgCdTe PICNIC infrared focal plane array (Vural
et al. 1999).  The PICNIC array is a hybrid device with four
independent quadrant outputs.  This hybrid device is similar to
Rockwell's familiar NICMOS3 array; it has identical unit cell size,
number of outputs and general architecture, and uses a modified
multiplexer design and better fabrication techniques to improve the
noise and amplifier glow.  The PICNIC array detector material is
HgCdTe with a band-gap corresponding to 2.5 $\mu$m.  The requirement
for moderate dark current ($< 10^2$ e$^-$ s$^{-1}$) necessitates
operation below 100~K. A cold environment also is required to minimize
the thermal radiation from room-temperature surfaces reaching the
detector. A simple calculation shows that formation of a cold pupil to
match the acceptance F-cone of the camera to the telescope is
necessary if useful operation in $K$-band is expected from the
instrument. Cold baffles alone provide an ineffective and impractical
method of controlling this thermal background.

\subsection{Optical design}

The camera is designed to operate at the Cassegrain focus of the
30-inch telescope of Leuschner Observatory, which is located 10 miles
east of the Berkeley Campus in Lafayette.  The 30-inch is a
Ritchey-Chretien with a nominal F/8 final focal ratio, yielding a
scale of 33.8 arc seconds per mm. The seeing at Leuschner is typically
2-3 arc seconds FWHM at visible wavelengths. At this scale the 40
$\mu$m PICNIC pixels project to 1.35 arc seconds on the sky.

Direct imaging is a simple yet natural solution which takes advantage
of the large corrected field delivered by the Ritchey-Chretien design.
Unfortunately, direct imaging is an unacceptable solution because of
the detector's sensitivity to thermal radiation.  Consider installing
the PICNIC array in a vacuum vessel typical of CCD Dewars.  Suppose,
optimistically, that the cold, dark interior of the Dewar restricts
the solid angle of ambient radiation illuminating the detector to an
F/3 cone.  The black-body radiation from the instrument flange and
primary mirror support structure at 293 K in this solid angle in an
$H$-band filter (1.5-1.8 $\mu$m) is 2500 photons s$^{-1}$ per
pixel. This should be compared with the sky signal of 6000 s$^{-1}$
per pixel (for $H$ = 14 mag. per square arc second) and the detector dark
current of a few electrons per second per pixel.  At $K_s$ (2.0-2.3
$\mu$m) the unwanted thermal radiation is two orders of magnitude
greater ($4.5\times 10^5$ photons s$^{-1}$ per pixel) than it is at
$H$.  In Fall 1999 we deployed a prototype camera to test the array
read-out electronics, vacuum Dewar, and the closed cycle He
refrigerator.  To speed the development we eliminated all optical
components apart from a fixed filter that was located directly in
front of the focal plane array.  Stray radiation was controlled only
by a cold baffle tube.  This configuration, as predicted, suffered
from high background to the extent that it was unusable in the
$K_s$ band
because the detector saturated in the minimum integration time
(0.66~s). Both theory and practice show that formation of a cold pupil
within the Dewar is an essential to successful IR operation.

\begin{figure}[h]
\epsfig{figure=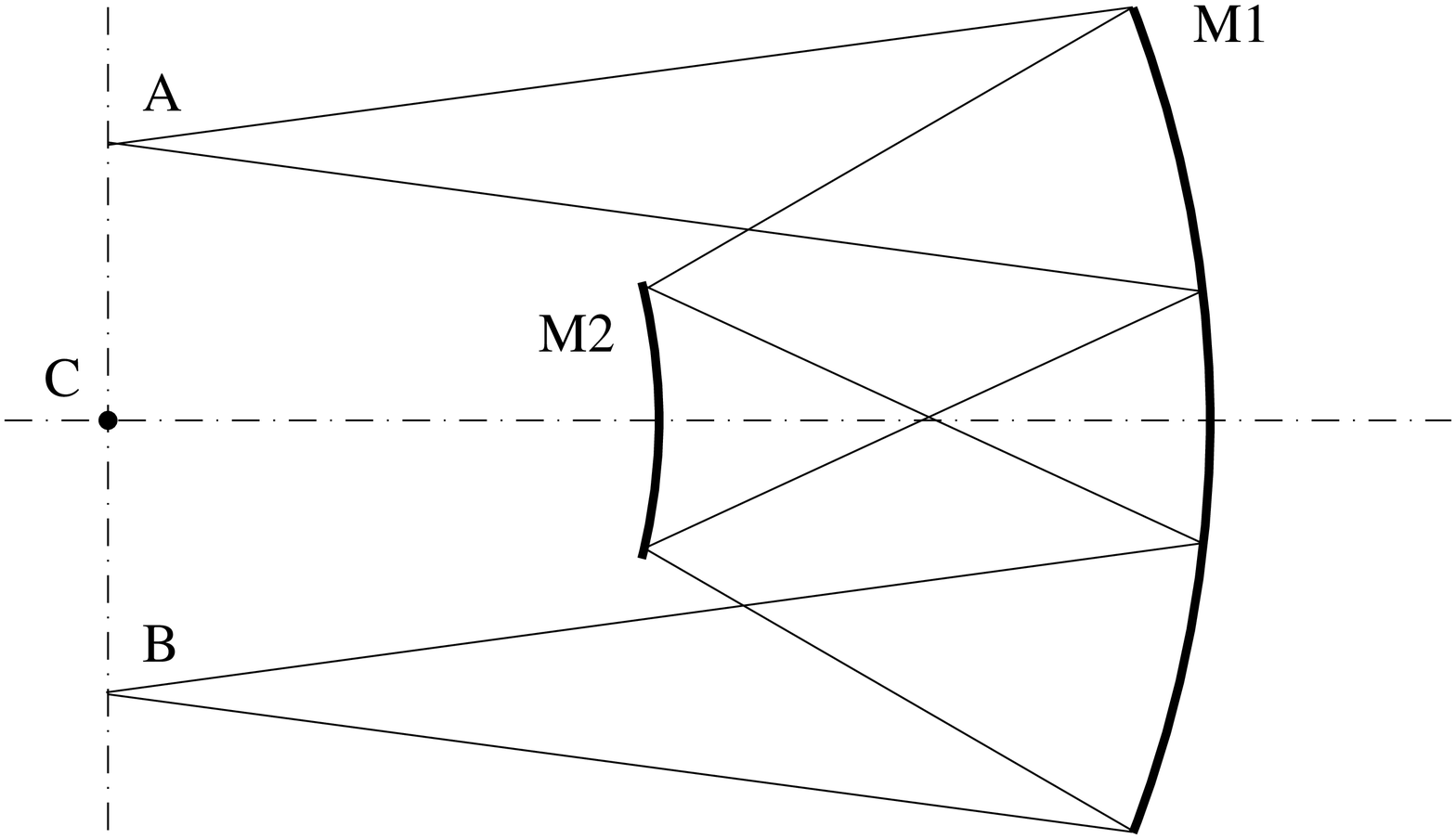,width=2.75in,angle=0}
\caption{\small The off-axis Offner system relays the image at $A$ to $B$
with unit magnification.  The concave mirror $M_1$ and the convex
mirror $M_2$ are concentric about the point $C$. $M_2$ is placed half
way between $M_1$ and the focal plane $AB$ to give a zero
Petzval sum, i.e., $R_1 = -2R_2$. \label{offnerlinedrawing}}
\end{figure}

An infrared camera, known as KCAM, which was built with similar
instructional objectives, is in use at UCLA (Nelson et al. 1997).
This camera uses a ZnSe singlet to relay the image with
unit-magnification from a 24-inch F/16 telescope onto a NICMOS-3 IR
focal plane array.  Simultaneously this lens forms a real image of the
telescope pupil where stray light can be controlled by a cold stop.  A
refractive design combines the advantages of simple assembly and
alignment. The high refractive index of ZnSe ($n = 2.43$ at 2 $\mu$m)
permits effective control of spherical aberration, while the relative
slowness of the UCLA system means that off-axis aberrations are
negligible.  ZnSe is moderately dispersive, which means that
refocusing between $J$, $H$, and $K$ is necessary---this is not an
issue for KCAM which has a fixed filter. Our faster (F/8)
configuration together with our desire to eliminate the need for
refocusing forced us to consider doublet designs.  However, we were
unable to achieve good correction for geometric and chromatic
aberrations over a 6 arc minute field without resorting to custom
lenses, exotic salts or glasses, and expensive anti-reflection
coatings (the reflection loss for uncoated ZnSe is 32\%.)

\begin{figure}[htb]
\begin{center}
\epsfig{figure=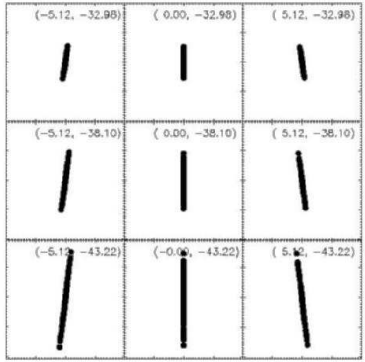,width=2.in,angle=0}
\end{center}
\caption{\small 
Spot diagrams for an F/8 Offner relay.  The nine panels correspond to
the center, edges and corners of the PICNIC array. Each box is the
size of a single PICNIC pixel (40 $\mu$m) and is labeled with its
coordinates relative to the center of curvature of $M_1$ and $M_2$
(point $C$ in Fig. \ref{offnerlinedrawing}.)
Dimensions are in mm. The geometric image size is
limited by fifth-order astigmatism. \label{classical_offner}}
\end{figure}

We therefore decided to evaluate the potential of a reflective
design. The telecentric, unit magnification Offner relay (Offner 1975)
has been used with success in several infrared instruments (e.g.,
Murphy et al. 1995).  The Offner relay consists of two concentric
spherical mirrors and is free of all third-order aberrations.  This
geometry has two real conjugates in the plane which passes through the
common center of curvature and is perpendicular to the axis joining
the vertices of the two mirrors (Fig. \ref{offnerlinedrawing}).  The
first mirror, $M_1$, is concave while the secondary, $M_2$, is convex,
with a radius of curvature, $R_2 = -R_1/2$.  The Petzval sum of the
system is zero, because the combined power of the
two reflections at $M_1$ is
equal and opposite to that of $M_2$.  Because $M_1$ and $M_2$ are
concentric, an object placed at the common center of curvature will be
imaged onto itself with no spherical aberration.  Vignetting limits
the practical application of the Offner configuration on-axis. However,
vignetting drops to zero for off-axis distances equal to or greater
than the diameter of $M_2$.  Coma and distortion, aberrations that
depend on odd powers of the ray height, introduced by the first
reflection at $M_1$, are cancelled on the second reflection at this
surface because of the symmetry about the stop formed by $M_2$.
Fifth-order astigmatism limits the numerical aperture at which good
imagery is obtained.  $M_2$ is located at the mid-point between $M_1$
and its center of curvature; it therefore is located at an image of
the telescope pupil. This pupil image suffers spherical aberration.
Nonetheless, the image quality is good enough so that 
a cold stop at $M_2$ suppresses light that does
not emanate from the telescope pupil and unwanted thermal
emission from the telescope structure. The stop will also act as a
baffle to reduce scattered light at all wavelengths and should make it
easier to measure accurate flat fields.

\begin{figure}[htb]
\begin{center}
\epsfig{figure=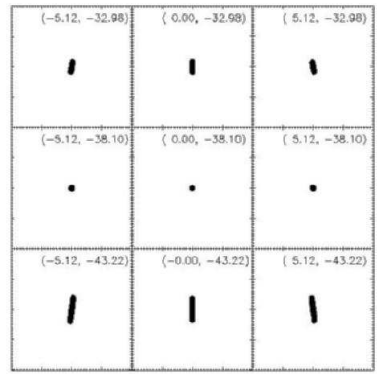,width=2.in,angle=0}
\end{center}
\caption{\small 
Spot diagrams for an F/8 corrected Offner relay. Layout is as in
Fig. \ref{classical_offner}.  Increasing the radius of $M_2$, while
keeping it concentric with $M_1$, introduces enough third-order
astigmatism to cancel fifth-order astigmatism at the central
field point.
\label{corrected_offner}}
\end{figure}

Fig. \ref{classical_offner} shows the imaging performance of a
classical F/8 Offner based on a 6-inch (152.4 mm) focal length
spherical primary (PN K32-836, Edmund Scientific) and a 76.2 mm
separation between the image and object.  Clearly the
performance is very good, with a central field point geometric size of
5 $\mu$m rms. The worst field point is the bottom right corner of the
focal plane (-5.12 mm, -43.22 mm) with 8 $\mu$m rms spot size.

\begin{figure}[htb]
\begin{center}
\epsfig{figure=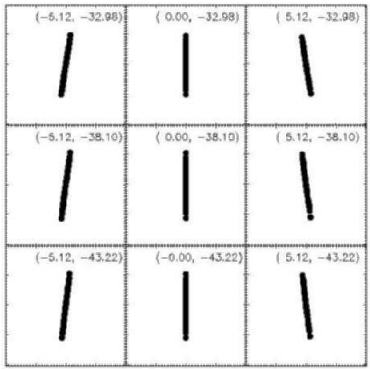,width=2.in,angle=0}
\end{center}
\caption{\small 
Spot diagrams for the modified F/8 Offner based on 6-inch (152.4 mm)
focal length spherical primary and a $R_2=155.04$~mm secondary.
Layout is as in Fig. \ref{classical_offner}.  This design has a slight
excess of third-order astigmatism, but perfectly adequate and uniform
performance of 5 $\mu$m rms spots over the entire field.
\label{actual_offner}}
\end{figure}

Our method for fabricating the secondary mirror is to select a
commercial plano-convex BK7 lens and aluminize its curved surface.
Choosing $M_2$ from a catalog makes it likely that exact achievement
of $R_2 = -R_1/2$ is not possible.  Selecting $M_2$ with a slightly
larger radius of curvature is advantageous.  Increasing the radius of
$M_2$, while keeping it concentric with $M_1$ introduces enough
third-order astigmatism to cancel fifth-order astigmatism at any given
field point. This cancellation, which occurs with a $R_2=153.6$~mm, is
shown in Fig. \ref{corrected_offner}.  The closest readily available
match plano-convex lens ($R_2=155.04$~mm; PN K45-284, Edmund
Scientific) has slight excess third-order astigmatism, but perfectly
adequate and uniform performance of 5 $\mu$m rms spots
(Fig. \ref{actual_offner}) over the entire field.

\begin{figure}[htb]
\begin{center}
\epsfig{figure=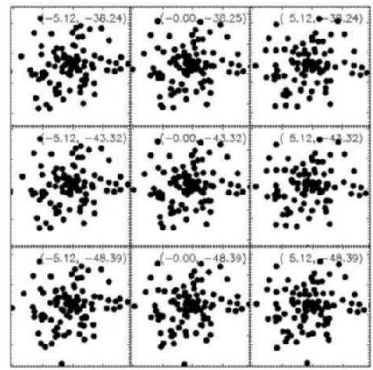,width=2.in,angle=0}
\end{center}
\caption{\small  The Offner design is robust against misalignments.  The spot
diagrams are for the modified F/8 Offner in Fig. \ref{actual_offner},
but with a 1 degree tilt of the secondary about its vertex. The rms
spot size is 11 $\mu$m. A tilt of this amplitude introduces 5.2 mm of
image motion which is easily detected. \label{offner_tilt}}
\end{figure}

\begin{figure}[thb]
\begin{center}
\epsfig{figure=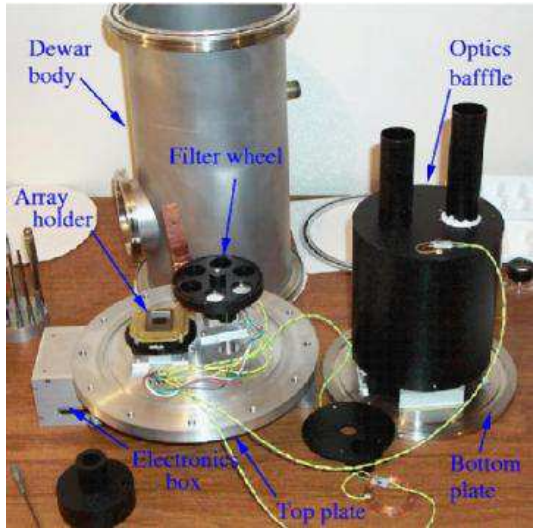,width=2.75in,angle=0}
\end{center}
\caption{\small The Dewar and internal components.  The Dewar consists of a
stainless steel cylindrical vacuum vessel and top and bottom plates.
The Dewar top plate carries the PICNIC array holder, the warm
electronics interface box, and the stepper motor/filter wheel
assembly. Both cold assemblies are attached to the top plate using
G-10 fiberglass tabs. The quartz window, and the hermetic MIL
connectors for array control, stepper motor, and temperature sensors
are also on this plate.  The bottom plate carries the primary mirror
cell (supported on G-10 tabs) and the optical baffle can. The main
baffle can also provides the mechanical support for the secondary and
doubles as a radiation shield. When assembled the exterior of optics
baffle can is wrapped with aluminized mylar.\label{components}}
\end{figure}

The alignment of the Offner relay presents no serious challenges
despite being an off-axis configuration. The performance is robust
against decenter of the secondary, which can be displaced by up to 4
mm before the geometric spot size exceeds one 40 $\mu$m pixel.
Secondary tilt is more critical; an error of 1 degree in tilt of $M_2$
about its vertex swells the size to an rms of 11 $\mu$m and the spots
fill one pixel (Fig. \ref{offner_tilt}).  However, a secondary tilt of
this amplitude also introduces 5.2 mm of image motion.  Optical
alignment, for example, can be achieved using a reflective reticle
placed at point $B$ in Fig. \ref{offnerlinedrawing} and an alignment
telescope which views point $A$.  In practice, two alignment jigs were
constructed which fit over the ends of the snouts on the optical
baffle can (see Fig. \ref{components}).  Each jig has a small central
hole which coincides with the focal plane.  One jig is illuminated
while its image is inspected at the other jig with a jeweller's loupe.
The tip and tilt of the primary mirror are then adjusted using three
alignment screws in the base of the mirror cell until the two jigs
appear coincident.

\subsection{Mechanical design}

\begin{figure}[thb]
\epsfig{figure=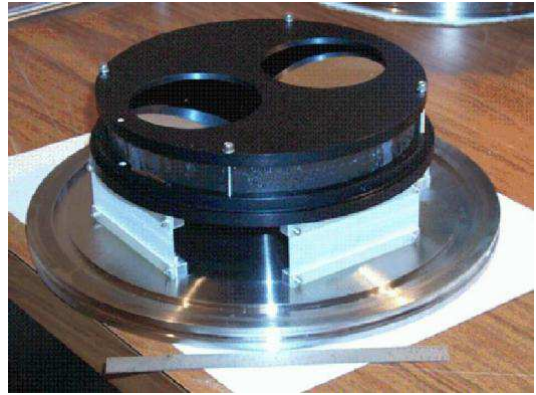,width=2.75in,angle=0}
\caption{\small The primary mirror of the Offner relay in its cell.  This F/1
spherical mirror is formed on a 6-inch diameter Pyrex 7740 blank.  The
mirror cell is attached to the Dewar bottom plate by four G-10 tabs
which provide a rigid, low thermal conductivity link to the Dewar
bottom. Three screws at the base of the mirror cell 
provide tip-tilt adjustment. 
\label{mirrorholder}}
\end{figure}

\begin{figure}[thb]
\epsfig{figure=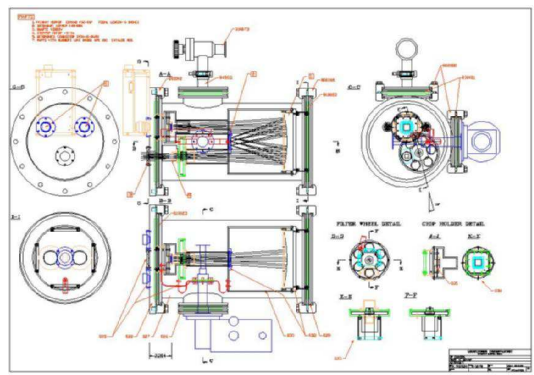,width=2.75in,angle=0}
\caption{\small A drawing of the infrared camera
Dewar, optics, cold head, and mechanisms. \label{ircameradrawing}}
\end{figure}

\begin{figure}[thb]
\epsfig{figure=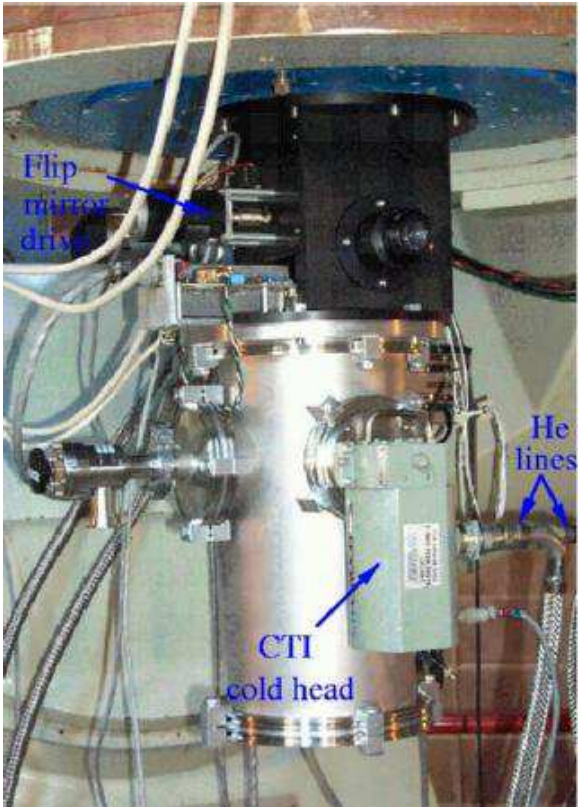,width=3.0in,angle=0}
\caption{\small 
The infrared camera mounted
at the Cassegrain focus of
the 30-inch telescope at Leuschner Observatory.
The mechanical interface to the Cassegrain flange is
via a mounting box which accommodates a 
computer-controlled flip
mirror. The flip mirror directs the light
to an eyepiece or a CCD camera; when swung out
of the way the infrared camera views the 
secondary directly. \label{ontelescope}}
\end{figure}

An overview of the layout of the Dewar is shown in
Fig. \ref{ircameradrawing} and an picture of the camera on the
telescope is shown in Fig. \ref{ontelescope}.  The Dewar body is a
semi-custom 8-inch diameter $\times$ 15-inch long cylindrical 304
stainless steel vacuum vessel (MDC Vacuum Products Corp., Hayward, CA)
with two 4-inch ports.  Both ends of the cylinder and the two 4-inch
ports are fitted with ISO LF flange fittings. The top plate
accommodates the quartz Dewar window and two hermetic MIL connectors
(Deteronics, South El Monte, CA); one for array control and read-out
and one for the stepper motor and temperature sensors. The chip
carrier and stepper motor/filter wheel assembly are anchored to the
top-end plate by G-10 fiberglass tabs (Fig \ref{components}).  G-10
provides an extremely stiff and low thermal conductivity (0.53 W
m$^{-1}$ K$^{-1}$) support.  The bottom plate provides the support for
the primary mirror of the Offner relay ($M_1$). Mechanical rigidity
for the primary mirror cell is provided by four fiberglass tabs
(Fig. \ref{mirrorholder}).  A cylindrical optical baffle can (Fig
\ref{components}) doubles as an radiation shield with input and output
snouts that mate with corresponding tubes on the filter wheel assembly
and the focal plane array housing (see Fig. \ref{ircameradrawing}).
The surfaces of the optical baffle can and other components adjacent
to the optical beam, such as the filter wheel and array housing, are
anodized black to suppress internal reflections. The outer surface of
the optical baffle can is wrapped with a single layer of aluminized
mylar to enhance its performance as a radiation shield.  The optical
baffle can attaches to the base of the mirror cell, and thereby
provides the alignment and mechanical support for the secondary
mirror. One of the two 4-inch flanges is used as an access port for
the model 21 CTI He refrigerator, while the other carries a vacuum
valve and pumping manifold. During final assembly this port also
provides access to the interior of the camera.

\subsubsection{Cryogenic filter wheel}

The camera is equipped with a cold filter wheel that can hold up to
six one-inch filters. The moment of inertia of the filter wheel is low
and the required precision is undemanding so the filter wheel is
driven directly by a stepper motor with no gear reduction.

Cryogenic mechanisms are frequently the most unreliable part of
infrared instruments. These mechanisms are usually actuated by a warm
external motor via ferro-fluidic feed-throughs.  A cryogenic motor
simplifies design and assembly of the Dewar by eliminating the
feed-through. Commercial vacuum-cryogenic stepper motors are
available, but they are more than an order of magnitude more expensive
than conventional motors.

We have followed the approach developed for the GEMINI twin-channel IR
camera at Lick Observatory which uses regular stepper motors with
specially prepared dry-lubricated bearings (McLean et al. 1994).  The
selected motor (PN Vexta PX244-03AA, Oriental Motor, Torrance, CA) is
two-phase, 1.8 degree per step, and rated 12VDC at 0.4A.  However, we
have found that disassembly and degreasing of the motor bearings in
isopropyl alcohol, omitting the time consuming step of burnishing with
MoS$_2$, is satisfactory. The motor home position is defined optically
using an LED and photodiode.

\subsubsection{Cryogenic design}

A typical $H$-band sky brightness of 14 mag per sq arc second yields about
6000 photons per second per 40 $\mu$m pixel on the 30-inch telescope.
Since we would like to retain the option of using narrow-band (1\%)
filters for imaging emission lines such as H$_2$ 1-0S(1) or Br$\gamma$,
our goal is to ensure that the thermal emission from the Offner relay
optics and the detector dark current is kept well below this level.
Thus, if feasible we would like to reduce any non-astronomical signal
to $< 10^2$ photons s$^{-1}$ pix$^{-1}$. Cooling the Offner optics by
70 K from 290 K is sufficient to reduce the $\lambda < 2.5\ \mu$m
thermal emission by three orders of magnitude, so that the background
level is reduced to a few hundred photons per second per
pixel. Reducing the dark current to 100 e$^-$ s$^{-1}$ requires
cooling the detector below 100~K.

The radiation load from the 290~K inner walls of the vacuum vessel
amounts to 5~W (for clean stainless steel walls with an emissivity of
0.05). Multistage thermoelectric coolers can maintain temperature
differences between hot and cold sides of as much as $150$~K, but
their efficiency is typically $< 10^{-3}$. Although thermoelectric
coolers are practical for CCD cameras which operate at smaller
temperature differences, they are not suitable for this application.
Liquid nitrogen is readily available; its 77~K boiling point and large
latent heat of vaporization (161 J ml$^{-1}$) render it an almost
ideal solution.  A cryogen vessel containing, say, 5 liters would be a
practical volume, given our design. However, this would require
filling once every two days. Installation of a low-emissivity floating
radiation shield could increase this interval by a factor of two.
Even with this extended lifetime the use of liquid cryogen is
incompatible with our objective of operating the infrared camera
remotely from Berkeley.  Our dilemma was solved when a Model 21 closed
cycle helium refrigerator (CTI Corp.)  was salvaged from the Berkeley
Radio Astronomy Lab's 85-ft telescope.  These cold heads are used
extensively by the semiconductor fabrication industry in cryopumps and
are readily available as second hand or refurbished items. The high
pressure He gas for the CTI cold head is supplied by a compressor
manufactured by Austin Scientific, Austin, TX.
We use only the 77~K station for cooling; the
20~K station has a vessel containing activated
charcoal embedded in epoxy for cryopumping. 

On cool down from room temperature the PICNIC array reaches its
initial operating temperature of 100~K after about three hours.  At
this temperature most of the functions of the camera can be checked
out.  After 12 hours the temperature of the optics and detector are
sufficiently stable for reliable operation.  In this state the 77~K
stage of the Model 21 cold head is at 66~K, the focal plane array is
at 82~K, the optics are at 155~K, and the filter wheel is at 187~K.
Over the course of a night the temperature stability of the array is
80~mK rms.  The Dewar temperatures, including the focal plane array,
show slow diurnal variations with an amplitude of 1-2~K which are
driven by variations in the ambient temperature.  These fluctuations
will cause significant dark current variations from night to night. To
guard against this the detector block is equipped with a heater
resistor that can be used in a temperature control loop.  To date,
nightly dark frames have proven adequate and the temperature control
servo has not be exercised.  The closed-cycle refrigerator has one
additional advantage related to the fact that the refrigerator can run
unattended for many weeks; the delicate detector and optical have
suffered fewer thermal cycles, and consequently less thermally induced
stress due to CTE mis-matches, than in a conventional liquid cryogen
system.

\subsection{Array control and readout}
\label{arraycontrol}

The PICNIC array is read out by clocking its multiplexer so that each
pixel is sequentially connected to one of the four output amplifiers.
When the pixel is accessed the charge is measured but not removed---a
reset line must be clocked to reset the pixel and clear the
accumulated charge.  Consequently, the charge can be read multiple
times to eliminate $kTC$ noise and 
improve the precision of the measurement.  Since the camera
does not have a mechanical shutter, timed exposures are made by
resetting the chip, waiting for the desired exposure time, then
reading the chip again. When no exposure is underway the chip is
constantly clocked with a reset waveform to flush out the accumulation
of charge.

\begin{figure}[thb]
\epsfig{figure=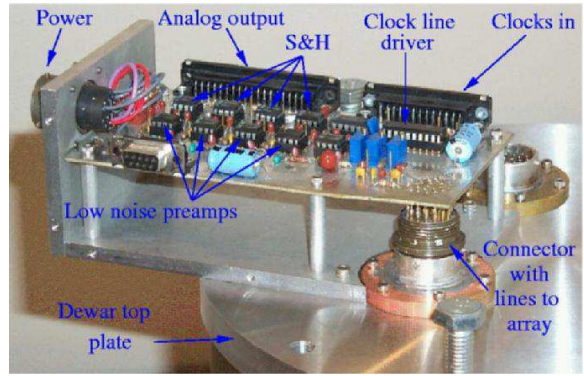,width=3.0in,angle=0}
\caption{\small The warm electronics interface box mounts on the top plate of
the Dewar. The array lines for clocks, biases, and quadrant outputs
pass through a MIL-style connector which is mounted directly on the
circuit card.  The clocks from the AWFG card (\S \ref{arraycontrol})
are buffered through a
CMOS line driver. Each quadrant's output channel has a low noise JFET
pre-amp (LF 356) and a sample and hold (S\&H) circuit (LF 398).  
\label{electronics}}
\end{figure}

The camera is controlled by commercially available cards that are
installed in a Intel/Pentium-based computer running Linux 2.0.34 which
is mounted on the side of the telescope about 1 meter from the
camera. This machine communicates to the outside world strictly
through ethernet.  A level shifter and interconnect box is mounted on
the vacuum Dewar (Fig. \ref{electronics}).

The PICNIC array requires six CMOS-level clocks and two 5 V power
supplies (one analog and one digital).  Each quadrant has two shift
registers for addressing pixels in the array---one horizontal register
and one vertical register. Each register requires two clocks. To obtain
a raster scan output, the horizontal register is clocked in the fast
direction and the vertical register is clocked in the slow direction.
PIXEL and LSYNC are the two required clocks for the horizontal
register. The PIXEL input clock is a dual edge triggered clock that
will increment the selected column on both edges (odd columns selected
on positive edges and even columns selected on negative edges). The
LSYNC clock is an active low input clock which will set a ``0'' in the
first latch and a ``1'' in the remaining latches of the shift
register, thereby initializing the shift register to select the first
column in the quadrant. Since this is asynchronous to the PIXEL clock,
LSYNC should be pulsed low prior to initiation of the first PIXEL
clock edge. The horizontal register selects which column bus will be
connected to the output amplifier.

These clock sequences are generated by an arbitrary waveform generator
card (PN PCIP-AWFG, Keithley Instruments). This card clocks out an
8-bit wide TTL-compatible word with a maximum pattern length of
32K. These words are converted to the 0-5 V levels suitable for
driving the PICNIC readout multiplexer and reset lines by a high speed
CMOS buffer/line driver (74HCT244). The four detector outputs (one for
each 128 $\times$ 128 quadrant) pass through warm monolithic
JFET-input operational amplifiers (PN LF356, National Semiconductor)
with a gain of 5.1 and a manually adjustable offset, a monolithic
sample and hold circuit (PN LF398, National Semiconductor) to a 16 bit
analog-to-digital converter card (PN DAS1602/16, Computerboards).  The
electronics interface box is shown in Fig. \ref{electronics}.  This
ADC has a 10 $\mu$s conversion time so the entire chip can be read in
660 ms.  The data from this card pass through a very large FIFO (Mega
Fifo, Computerboards) so that the data do not have to be read in real
time.

The array controller supports several array operations: 1)
read---reads chip with no reset; 2) reset---resets the chip but does
not read; 3) frame---reset, expose, read; 4) correlated double
sampling---resets and read, expose, read; 5) Fowler sampling---resets,
read, expose, read.  Here expose means wait the requested duration.
When a sequence involves two reads the difference of the second minus
first read is returned. This usually leads to a positive value of the
signal.

\subsubsection{Array clocking sequence generation}

Generating the clock waveforms for the PICNIC array is central to the
entire operation. The key problem is generating the sequence of bytes
that is to be down-loaded into the AWFG card. 
The six PICNIC clocks as well as the analog to digital
converter trigger clock have been assigned to the two 4-bit outputs of
the AWFG. The AWFG is capable of storing a waveform of up to 32K steps
long. When executing a waveform you must specify:
\begin{itemize}
\item length of waveform; 
\item number of times to execute the waveform;
\item clock divider specifying number of microseconds per step.
\end{itemize}

Unfortunately, the maximum waveform length of 32K is not large enough
to readout the entire PICNIC array.  Instead, we must specify a series
of waveforms and execute them sequentially. The AWFG board does allow
a new sequence to be loaded as the last sequence is finishing, if the
length of the sequence is the same and the clock divider does not
change.  We have chosen a waveform of length of 4128 clock steps. 

The basic step time is created from a 5 MHz clock divided by 25,
yielding a 5 $\mu$s base step time.  The full readout sequence is
created by stringing together multiple waveforms.  Various read modes
are supported. For example, the {\tt read} sequence which just reads
through the entire chip and digitizes the result is the combination of
two waveforms.  The {\tt frame} sequence is more complex and is
composed of three parts: 1) the chip is reset; 2) a variable length
wait time (the exposure); 3) The {\tt read} sequence.  The correlated
double sampling sequence (CDS) is similar, except that the software
subtracts the first values that are read just after the reset from the
second reads after the exposure is done and reports the difference.

The software to control the camera was written in {\tt C++} for a
client-server architecture. Linux device drivers for the three PC
cards were written in house.  The camera server communicates via
TCP/IP sockets and has commands to
\begin{itemize}
\item read camera temperatures and status; 
\item set readout waveforms;
\item read the PICNIC array.
\end{itemize}

The client program which is usually run on a remote machine (e.g., on
a workstation in the Berkeley undergraduate lab) reads the array and
current telescope information and creates a FITS file with data and
detailed header.  Since the observatory operations are automated,
scripts can be written to combine image acquisition and telescope
motions. This is vital in infrared observations where images are often
built up from multiple jogged exposures.

The minimum readout time of 660 ms is determined by the speed of the
analog to digital converter card.  Newer cards with 1~$\mu$s
conversion time are now available that would speed up the read out
time by a factor of ten with little change in the software.

\section{Performance}

The camera operated successfully during the Fall 2000 semester and
sufficient observing time has been accumulated by a class of 15
students to demonstrate satisfactory performance. The design has
proven reliable. For example, the Linux-based control software has
never crashed. The only significant problem was caused by the
commercial helium compressor.  During early operation the model 21 CTI
refrigerator delivered only marginal cooling capacity, with the PICNIC
array reaching only 100~K.  Intially, we suspected that the cold
head was defective since it was a salvaged unit. However, the helium
compressor failed catastrophically after only a few thousand hours of
operation; when it was replaced with a new unit the expected cryogenic
performance was achieved.

The adoption of a specially prepared motor for cryogenic operation was
a risk which allowed us to eliminate vacuum feedthroughs.  The stepper
motor has worked flawlessly, and this elegant solution for Dewar
mechanisms has paid off.  When the motor is driven continuously, or
holding current is applied, sufficient heat is deposited within the
Dewar so that after several hours of operation the temperature of the
optics and the detector rises by 10--15~K.  In normal operation, where
the filter wheel is active for a few seconds and then turned off,
this heat source is insignificant. Application of holding current is
unnecessary because the magnetic forces due to the permanent poles in
the motor are sufficient to keep the filter-wheel from moving.

Inspection of images delivered by the Offner relay at room temperature
leads us to suspect that it is essentially perfect, but it has not
proven possible to quantify the optical performance because the 
image blur has been dominated by seeing and telescope aberrations.
These aberrations are due to telescope collimation errors which are
apparent in direct CCD images.  The best image sizes delivered by the
camera have size of less than 2 pixels FWHM (2.6 arcseconds). However,
this is only achieved after careful focusing of the 
telescope secondary mirror.

\begin{figure}[htb]
\epsfig{figure=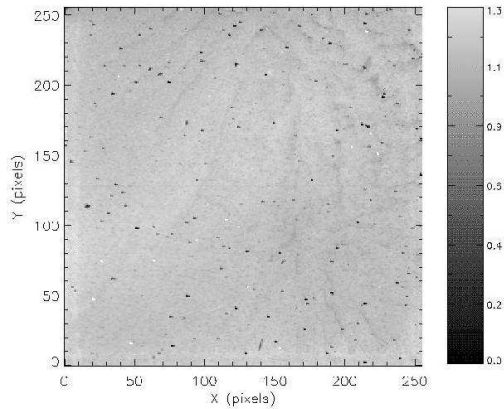,width=2.75in,angle=0}
\caption{\small A grey scale representation of the
$K_s$-band flat field. Bad pixels appear as black. 
The cosmetic properties of this array are very
good. See Fig. \ref{flatfieldhistogram} for
the pixel response histogram. The flat has has been
normalized so that the median value is unity.
\label{flatfield}}
\end{figure}

In the next three subsections we describe several characteristics of
the infrared array and the camera.  Exploration and quantification of
the camera performance forms the basis of the first lab performed by
students. Once they have understood the operation of the camera and
mastered the calibration of images they are ready to make astronomical
observations such as the one outlined in \S \ref{example-observation}.

\subsection{Array uniformity} 

\begin{figure}[thb]
\epsfig{figure=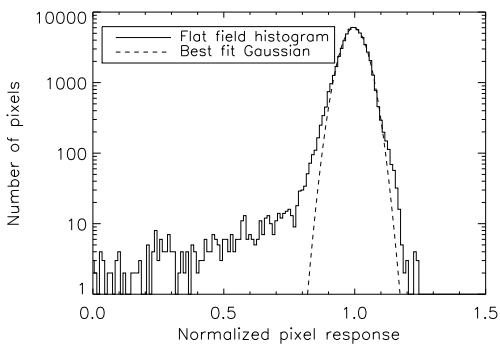,width=2.75in,angle=0}
\caption{\small The histogram of the normalized pixel
response of a $K_s$-band flat. 
The rms of the best fit Gaussian 
distribution is 0.04 and 
3\% of the pixels are ``cold'', forming a tail 
with poor response.
\label{flatfieldhistogram}}
\end{figure}

Under high-background conditions flat field accuracy is most important
systematic error which limits sensitivity of an array detector to
faint sources.  Modern arrays deliver very uniform pixel-to-pixel
response which means that students can easily construct a flat field
that is adequate for all practical purposes. Even this array, which is
nominally designated an engineering grade device, has excellent
cosmetic characteristics (see Figs. \ref{flatfield} \&
\ref{flatfieldhistogram}).  This is illustrated qualitatively by
Fig. \ref{flatfield} which shows a grey-scale image of a $K_s$-band
flat field. This flat is a weighted mean of 11 dark-subtracted
twilight sky images.  Taking twilight sky images requires careful
timing; if they are taken too early the detector saturates, taken too
late then there is insufficient signal to form a high signal-to-noise
flat.  For our pixel etendue ($1.6 \times 10^{-7}$ cm$^2$ str) the
optimum time for taking $K_s$-band images is between sunset or sunrise
and 6-degree twilight.  A field at high galactic latitude is suitable
to avoid stars.  The telescope is jogged between each exposure so that
stars occupy different pixels in subsequent pictures.  Any pixels
containing obvious stars are given zero weight and the frames are
combined into a weighted mean where outlying pixels (faint stars or
cosmic rays) are rejected if their deviation from the mean exceeds
three sigma.  In a flat normalized so that the median pixel value is
1.0 the best fit Gaussian has an rms of 0.04
(Fig. \ref{flatfieldhistogram}).  Three percent of the pixels fall in
a tail with low response outside of this distribution, and the array
can be considered 97\% operable.

\subsection{System noise}

Under the assumption that the variance in the camera signal is the sum
of a constant read noise and Poisson fluctuations it is
straightforward to measure the read noise and gain.  The data required
for this measurement are a sequence of exposures of increasing
integration time while the camera views a source of spatially uniform
illumination.  A pair of exposures is acquired at each integration
time.  The mean signal level was measured by subtracting a bias frame
and the variance computed by differencing the two frames---since the
uniformity of flat field is very high the frames were not flat
fielded.  A straight-line, least squares fit to the linear
mean-variance relation yields the readout noise as the intercept and
the gain as the slope.  The results of this analysis yield 30 e$^-$
per data number (DN), and a read noise of 70 e$^-$ rms. The majority of
this is detector noise, since if one grounds the input to the data
acquisition system the variation is on the order of 30 e$^-$ rms.  A
science grade PICNIC array is expected to have a read noise of about
10 e$^-$ rms. Despite this elevated detector noise, the data are
background limited in all but the shortest exposures. The
array saturates at 22,000 DN (660,000 e$^-$) and shows
$<$ 1\% nonlinearity at up to 80\% of this level. 

\begin{table}[htb]
\begin{center}
\caption{Camera Zero Points \label{zeropoints}}
\vskip 0.15 in
\begin{tabular}{ll}
\hline
\hline
Band & $m_{zpt}$ \\
\hline
$J$   & 18.0 \\
$H$   & 18.2 \\
$K_s$ & 17.5 \\
\hline
\end{tabular}
\end{center}
\end{table}

\subsection{Throughput}

The efficiency of the camera is expected to be high. The detector
quantum efficiency is 60-70\% over most of the operating wavelength
range.  The average filter transmissions are 86\% ($J$), 83\% ($H$),
and 92\% ($K_s$) between the half-power points. The protected Al
coating for the primary, $M_1$, has an infrared reflectivity of 97\%,
and the reflectivity of bare Al on the secondary, $M_2$, is probably
similar. The quartz Dewar window has a transmission of 94\%, giving a
predicted camera efficiency of 50\%. The system efficiency,
including the telescope and atmosphere,
determined from observation of standard
stars is 30\% ($J$), 39\% ($H$), and 44\% ($K_s$). 
The camera zero
points in magnitudes on a Vega based scale, 
defined as
\begin{displaymath} 
m_{zpt}  =  m_{star} + 2.5 \log_{10} ({\rm DN~s^{-1} } )
\end{displaymath}
are listed in Table \ref{zeropoints}, where DN~s$^{-1}$ is the count
rate in for a star of magnitude $m_{star}$.

\begin{figure}[thb]
\epsfig{figure=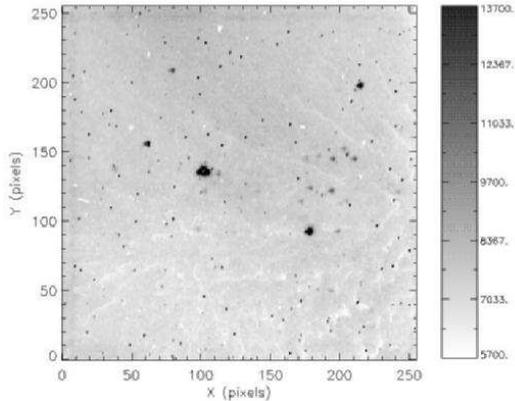,width=2.75in,angle=0}
\caption{\small A single raw $K_s$-band frame of
NGC~2024. This is a 10 second exposure.
Only the
brightest stars ($K \simeq 8$~mag) can be seen in this image.
One pixel corresponds to 1.35 arc seconds and the field
of view is 5.8 $\times$ 5.8 arc minutes.
\label{oneKframe}}
\end{figure}

\subsection{An example observation}
\label{example-observation}

The region NGC~2024 (Orion B, W12) is a well-studied site of massive
star formation located at a distance of about 415~pc (Anthony-Twarog
1982).  It is also known as the Flame Nebula, because of the shadow of
prominent lanes of extinction formed against a background of H$\alpha$
emission from ionized gas.  The ionizing stars for this nebula are
hidden behind this dust lane.  Near-infrared radiation from these
stars penetrates the dust and a dense stellar cluster is revealed in
the optically-dark region separating the two halves of the
nebula. Observations of NGC~2024 provide a dramatic demonstration of
the ability of near-infrared radiation to penetrate dust.

\begin{figure}[thb]
\epsfig{figure=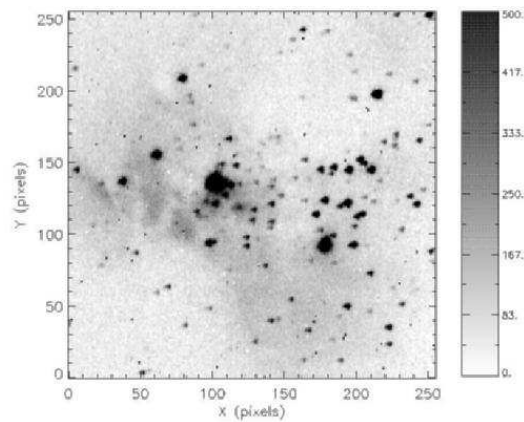,width=2.75in,angle=0}
\caption{\small Flat-fielding and sky subtraction of
the raw frame in Fig. \ref{oneKframe} reveals
significantly more detail, including the low-contrast 
nebulosity. The coma and astigmatism apparent in this
image are due to telescope colimation errors. 
\label{oneKframeffss}}
\end{figure}

NGC~2024 was observed on 2000 Nov 4. Eight 10 s exposures were taken
in $J$, $H$, and $K_s$. The telescope was rastered in a $3 \times 3$
(leaving out the central location) with offsets of 36 arc seconds in
RA and DEC between each pointing. Exposures were collected in $J$,
$H$, and $K_s$ for the target and then the telescope was offset to
blank sky 6 arc minutes to the east and the filter sequence was
repeated.

\begin{figure}[thb]
\epsfig{figure=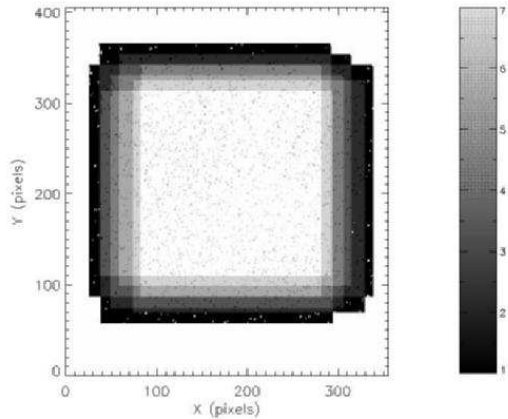,width=2.75in,angle=0}
\caption{\small The exposure mask for this observation.  
Seven good frames
were acquired, each one at a different telescope pointing. This image
represents the projection of good pixels onto the field of view.  Bad
pixels can identified as local minima in
the exposure mask.  Dithering the telescope 
ensures that every point in the field of view is covered by a
functioning pixel.
\label{expmask}}
\end{figure}

\begin{figure}[thb]
\epsfig{figure=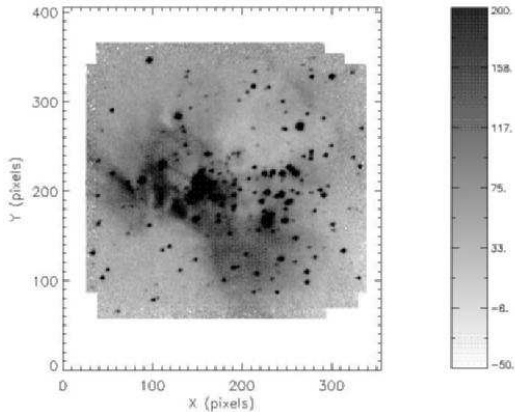,width=2.75in,angle=0}
\caption{\small The final $K_s$-band mosaic for NGC 2024. The total exposure
time in the center of the mosaic is 70 seconds.  Heavy foreground
extinction due to interstellar dust reddens many of the stars in this
cluster.  Even at this wavelength, where the extinction is one tenth
of that at visible wavelengths, prominent dark dust lanes are
apparent. \label{Kmosaic}}
\end{figure}

\begin{figure}[thb]
\epsfig{figure=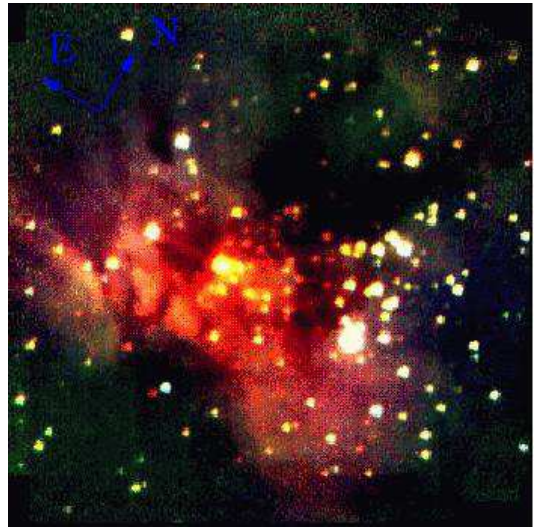,width=2.75in,angle=0}
\caption{\small A three-color infrared image of NGC 2024.  Red corresponds to
$K_s$, green is $H$, and blue is $J$.  The reddened stars at the
center of the image are responsible for ionizing the gas in this
region.  The mosaic has been cropped to a field of view of 5.8 arc
minutes $\times$ 5.8 arc minutes.  The exposure time is 80 seconds per
filter and the limiting magnitude is 14.2 at $J$ and 13 at
$K_s$. North and east are indicated.
\label{ngc2024color}}
\end{figure}

Fig. \ref{oneKframe} shows an individual $K_s$ band image. Only the
brightest stars ($K \simeq 8$~mag) can be seen in raw images. A
sky-subtracted and flat-fielded image (Fig. \ref{oneKframeffss}) shows
significantly more detail, including the nebulosity.  The exposure
mask for this observation is shown in Fig. \ref{expmask}, and the
final mosaic composed using using a weighted sum of the registered
images is shown in Fig. \ref{Kmosaic}. The $J$, $H$, and $K_s$ images
have been registered and combined into a three color composite. The
optically invisible, highly reddened, stellar cluster behind the
central dark lane stands out clearly (Fig. \ref{ngc2024color}).

\section{Conclusions}

We have designed and built a high performance near-infared array
camera for the 30-inch telescope at Leuschner Observatory. Our design
eliminates custom parts in favor of commercial components.  The array
control and readout electronics are composed entirely of standard
PC cards. The only custom board is the interface which buffers
the clocks from the waveform generator and amplifies the analog signal
from the detector outputs. The Offner-based optics relay the telescope
image at unit magnification via a cold stop which provides effective
control of thermal radiation from the telescope and its environment.
The camera has been used successfully in an undergraduate laboratory
class in Fall 2000.

\acknowledgments

Construction of this camera was made possible by the National Science
Foundation through Instructional Laboratory Initiative grant 9650176
and the generous support of Kadri Vural at the Rockwell Science
Center, Thousand Oaks, CA.  The Berkeley Astronomy Department provided
matching funds for this enterprise.  We thank Prof. I. S. McLean
(UCLA) for advice during the design and construction of the
camera.  W. T. Lum and Dr. D. Williams of Berkshire Technologies helped
with the Dewar and C. Chang (UCB, Radio Astronomy Lab) did the PC
layout.  Prof. C. Heiles provided invaluable guidance and
encouragement.




\clearpage








\clearpage


\end{document}